\documentclass[showpacs,preprintnumbers,amsmath,amssymb,prd,superscriptaddress,twocolumn]{revtex4-1}

\usepackage{txfonts}
\usepackage{graphicx}
\usepackage{dcolumn}
\usepackage{bm}
\usepackage{color}
\usepackage{subfigure}  
\def\comment#1{}

\def\slashchar#1{\setbox0=\hbox{$#1$}           
	\dimen0=\wd0                                 
	\setbox1=\hbox{/} \dimen1=\wd1               
	\ifdim\dimen0>\dimen1                        
	\rlap{\hbox to \dimen0{\hfil/\hfil}}      
	#1                                        
	\else                                        
	\rlap{\hbox to \dimen1{\hfil$#1$\hfil}}   
	/                                         
	\fi}                                         %

\def\nablab{{\mbox{\boldmath $\nabla$}}}

\newcommand{\blue}{\textcolor{black} }

\begin{document}

\title{\blue{Duality of a compact topological superconductor model and the Witten effect}}
\author{Flavio S. Nogueira}
\affiliation{Institute for Theoretical Solid State Physics, IFW Dresden, Helmholtzstr. 20, 01069 Dresden, Germany}
\affiliation{Institut f{\"u}r Theoretische Physik III, Ruhr-Universit\"at Bochum,
Universit\"atsstra\ss e 150, 44801 Bochum, Germany}

\author{Zohar Nussinov}
\affiliation{Physics Department, CB 1105, 
Washington University, 1 Brookings Drive, St. Louis, MO 63130-4899}

\author{Jeroen van den Brink}
\affiliation{Institute for Theoretical Solid State Physics, IFW Dresden, Helmholtzstr. 20, 01069 Dresden, Germany}
\affiliation{Institute for Theoretical Physics, TU Dresden, 01069 Dresden, Germany}

\date{Received \today}

\begin{abstract}

\blue{We consider  a compact abelian Higgs model  in 3+1 dimensions 
	with a topological axion term and construct its dual theories for both bulk and boundary at strong coupling. The model may be viewed as 
	describing a superconductor with magnetic monopoles, which can also be  
	interpreted as a field theory of a topological Mott insulator. We show that this model is dual to a non-compact topological 
	field theory of particles and vortices. It has exactly the same form of a model for superconducting cosmic strings with an axion term. 
	We consider the duality of the boundary field theory at strong coupling and show that in this case $\theta$ is quantized as $-8\pi n/m$ where 
	$n$ and $m$ are the quantum numbers associated to electric and magnetic charges.    	
These topological states lack a non-interacting equivalent. }

\end{abstract}

\maketitle

\section{Introduction}

A plethora of topological states of matter have been identified and classified during the past 
decade~\cite{Hasan-Kane-RMP,Zhang-RMP-2011,Classif}. These include material realizations such as strong topological insulators (STI). Interestingly, the microscopic electronic structure of these materials can be very different. However, some properties of an STI, set by topology, are universal. A celebrated example is the bulk-boundary correspondence guaranteeing the presence of surface states that are protected by the bulk topology.
Another incarnation of this universality arises in the field-theoretical description of the electromagnetic response of STIs: it is governed by the canonical Maxwell Lagrangian supplemented by a topological term - the axion or $\theta$-term, 
$\sim\theta~{\bf E}\cdot{\bf B}$, which quantizes the electromagnetic response~\cite{Qi-2008}. 

Instead of an STI we consider a compact abelian Higgs model in 3+1 dimensions  
with a $\theta$-term \cite{Ryu-2012,Qi-Witten-Zhang,Roy,Nogueira-Sudbo-Eremin}, 
which may be interpreted as an 
 effective field theory for a topological Mott insulator and show that it is dual to an axionic superconductor model \cite{Ryu-2012,Qi-Witten-Zhang,Roy,Nogueira-Sudbo-Eremin} 
 where both particle and 
 vortex degrees of freedom appear in the Lagrangian.  The Lagrangian of the dual theory is similar to the one of 
 a model for superconducting vortex strings \cite{Witten-SC-strings}, 
 except that it also features a $\theta$-term, which causes a topologically induced charge coupling for 
 the vortex lines.  Such an interacting field theory can be physically understood in terms of an experimental setup consisting of  
a superconducting slab sandwiched between two semi-infinite STIs (see Fig.~\ref{Fig:TI-SC-TI}). The $\theta$-term 
of the STI couples to the electrodynamics of vortex lines in the superconductor. This can be shown to lead to a charge fractionalization mechanism at the interfaces similar 
to the Witten effect, although no magnetic monopoles are present in this setting (see Section II). Thus, the Witten effect with 
charge fractionalization due to magnetic monopoles in 
the compact abelian Higgs model with an axion term maps via duality into a Witten effect associated to vortex lines.  

It is well-known that {\it without} the topological axion term the compact Maxwell theory in 3+1 dimensions exhibits a confinement-deconfinement transition \cite{Guth}. This transition can be understood by exploiting the duality of the compact Maxwell theory to the non-compact abelian Higgs model \cite{Peskin}. In the dual Higgs model vortex lines  correspond to worldlines of magnetic monopoles in the original model. Hence, the phase transition in the dual Higgs model corresponds to the confinement-deconfinement transition in the original compact U(1) Maxwell electrodynamics. The situation in 3+1 dimensions is 
quite different from the one in 2+1 dimensions where test charges are permanently confined \cite{Polyakov}, with the Wilson loop satisfying the 
area law. 
Indeed, it is well known that compact Maxwell theory 
in 2+1 dimensions is dual to a Coulomb gas of magnetic monopoles (actually in this case it is more technically correct to speak of instantons). The 
sine-Gordon Lagrangian yields an exact field theory representation of a Coulomb gas in any dimensions \cite{Froehlich-Spencer-1981}. 
In 2+1 dimensions the sine-Gordon 
theory is always gapped, so no phase transition occurs in this case \cite{Polyakov,Goepfer-Mack}.   

The duality transformation can also be carried out for the case of a compact abelian Higgs model. The exact result has been obtained 
for a model defined on a $d$-dimensional lattice long time ago \cite{Einhorn}. Generally,  when Higgs fields are included, the dual 
model is given by a vector or tensor Coulomb gas, depending on the dimensionality. In this paper we will find it useful to consider 
besides the complete duality transformation leading to a Coulomb gas, also a partial duality transformation, where the Higgs field 
is still present, while the magnetic monopole degrees of freedom are mapped on a dual Higgs sector representing the ensemble of 
world lines of magnetic monopoles as vortex lines.  The resulting model in 3+1 dimensions corresponds to one of superconducting 
vortex strings mentioned above. If the original Higgs and gauge fields are also integrated out, the field theory corresponding to 
the duality discussed on the lattice by Cardy \cite{Cardy-theta} and Cardy and Rabinovici \cite{Cardy-Rabinovici} is obtained.  

There is a question as to what happens with the duality at the boundary, which is important for topological states of matter. 
The compact Maxwell theory in 3+1 dimensions has a (2+1)-dimensional boundary. 
Thus, naively, we may think that the boundary theory is just compact Maxwell electrodynamics in 2+1 dimension. In such a case, the theory at the boundary 
will not exhibit a phase transition while the theory in the bulk will. However, this naive expectation clearly fails for 
the corresponding dual theory, since using the same logic we would expect that the dual model at the boundary is just the dimensionally 
reduced theory, i.e., the abelian Higgs model in 2+1 dimensions. This is obviously not the case, since the dual of compact Maxwell 
theory in 2+1 dimension is a sine-Gordon theory. Thus, the correct prescription to find the boundary dual theory is to dualize the dimensionally 
reduced model at the boundary. For the case of the axionic Higgs model we consider, the $\theta$-term generates a Chern-Simons term at 
the boundary.  We will show that in this case $\theta$ becomes fractionally quantized in the infinitely coupled regime. 

The plan of the paper is as follows. In Section II we discuss the Witten effect and derive a variant of it that also works with vortex lines. This 
result will serve to relate our duality to a physical problem of topological insulators coupled to type II superconductors \cite{Josephson-Witten}. In Section 
III, we introduce the compact axionic abelian Higgs model and show that it is equivalent to a non-compact model for superconducting 
vortex strings, thus establishing an exact mapping between a Higgs model containing monopoles and a model containing vortices and 
two Higgs fields.  In Section IV we discuss the duality transformation building on the results obtained in Section III. Section V discusses 
the boundary dual theory at strong coupling in the lattice.  
In Section VI we briefly comment on possible generalizations in the framework of quantum 
critical phenomena associated to the nonlinear $\sigma$ model. Section VII 
concludes the paper and in Appendix A we give further details on the calculations 
presented in the  main text.

\section{Witten effect in electrodynamics}

\subsection{Electromagnetic variant of the Witten effect with monopoles and vortex lines} 

\begin{figure}
	\centering
	\includegraphics[width=.9\columnwidth]{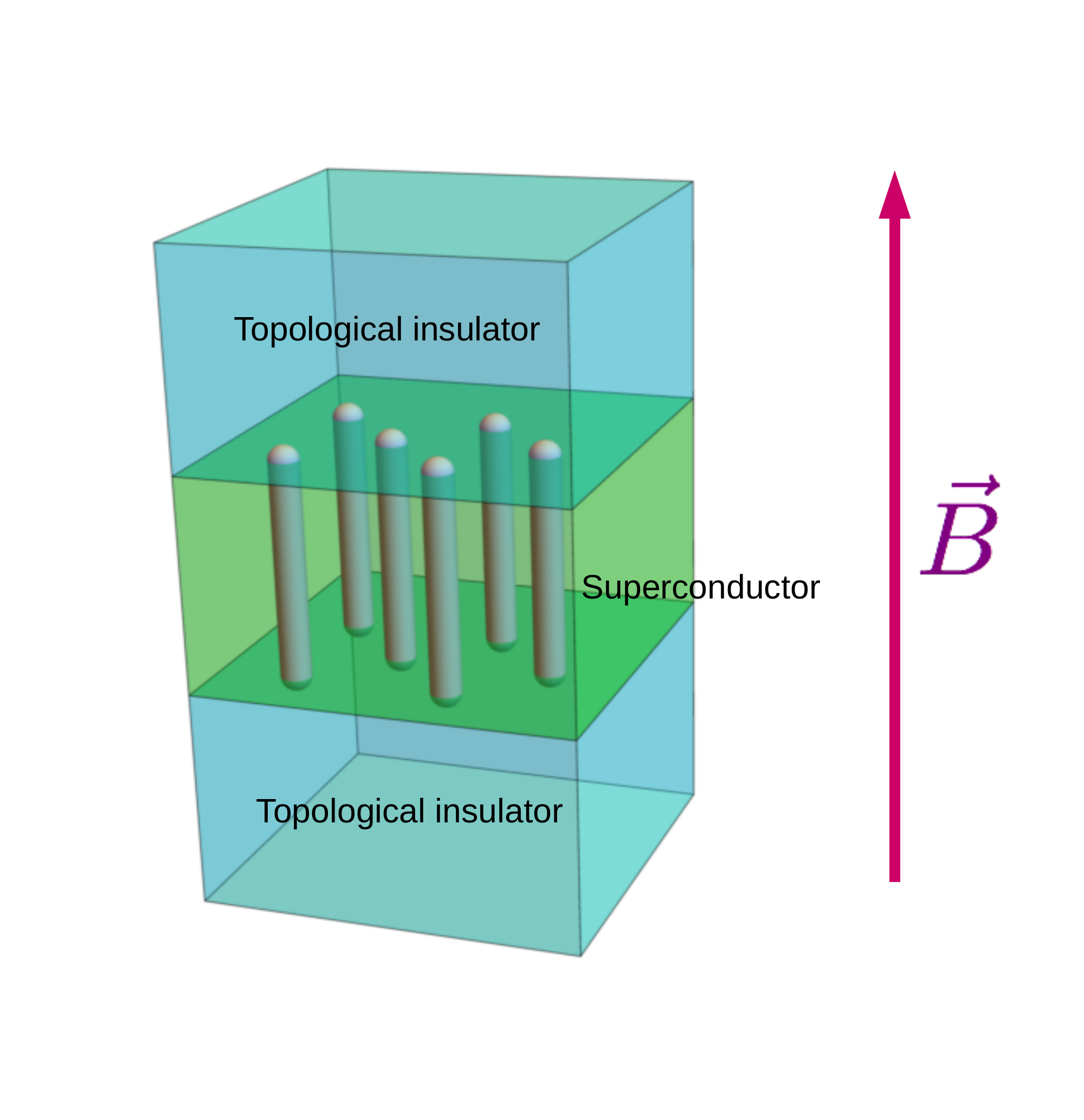}
	\caption{(Color online) Schematic view of a type II superconductor sandwiched between two STIs in presence of a magnetic field $\vec{B}$.  Due to the topological magnetoelectric effect, the vortex lines, represented by straight flux tubes, acquire an electric polarization. }
	\label{Fig:TI-SC-TI}
\end{figure}

The Lagrangian for electrodynamics with an axion term is given by
\begin{equation}
\label{Eq:ED-L}
{\cal L}=\frac{1}{8\pi}({\bf E}^2-{\bf B}^2)+\frac{e^2\theta}{4\pi^2}{\bf E}\cdot{\bf B}-\rho\phi-{\bf j}\cdot{\bf A}.
\end{equation}
The standard Maxwell equations are modified by the presence of the $\theta$ term. The new relations are easily obtained by computing the electric displacement vector ${\bf D}$ and the magnetizing field ${\bf H}$ via 
\begin{equation}
{\bf D}=\frac{\partial{\cal L}}{\partial {\bf E}},~~~~~~~{\bf H}=-\frac{\partial{\cal L}}{\partial {\bf B}},
\end{equation}
and inserting these results in the standard Maxwell equations.  The important equation for the Witten effect is the Gauss law, 
\begin{equation}
\label{Eq:Gauss}
\nablab\cdot{\bf E}=4\pi\rho-\frac{e^2}{\pi}\nablab\cdot(\theta {\bf B}). 
\end{equation}
Thus, unless magnetic monopoles are present, the Gauss law does not change if $\theta$ is uniform. 
Since $\nablab\cdot{\bf B}=\rho_m$, where $\rho_m$ is the magnetic monopole density, the integral form of the Gauss law 
reads
\begin{equation}
\label{Eq:Witten-general}
Q=q-\frac{e^2\theta}{4\pi^2}q_m-\frac{e^2}{4\pi^2}\int_V d^3r\nablab\theta\cdot{\bf B}.
\end{equation}
Here, $q_m$ is the magnetic charge, which fulfills the Dirac condition, $qq_m=2\pi$.   If $\theta$ is uniform 
and $q=ne$ (with integer $n$), Eq. (\ref{Eq:Witten-general}) yields the  charge fractionalization by monopoles of the Witten effect \cite{Witten-effect}, 
\begin{equation}
\label{Eq:Q-Witten}
Q=e\left(n-\frac{\theta}{2\pi}\right).
\end{equation}
In a condensed matter system we generally do not have intrinsic magnetic monopoles, but surface states provide yet another 
form of the Witten effect, due to the last term of Eq. (\ref{Eq:Witten-general}). Indeed, although in STIs $\theta$ is uniform, the presence of a surface leads to a nonzero value for the integral in Eq. (\ref{Eq:Witten-general}). Thus, if $\theta$ has a uniform 
value for  surfaces at $z=0$ and $z=L$, and ${\bf B}=B(r)\hat {\bf z}$  depends only on the radial coordinate $r$, we will obtain after setting $q_m=0$, 
\begin{eqnarray}
\label{Eq:Witten-general-1}
Q&=&q-\frac{e^2}{4\pi^2}\int_0^Ldz \frac{d\theta}{dz}\int d^2rB(r)
\nonumber\\
&=&q-\frac{e^2}{4\pi^2}[\theta(L)-\theta(0)]\Phi_B.
\end{eqnarray} 
The above constitutes a variant of the Witten effect when the magnetic flux $\Phi_B$ is nonzero. Fig. \ref{Fig:TI-SC-TI} illustrates a physical 
situation where Eq. (\ref{Eq:Witten-general-1}) is realized, with a type II superconductor is 
sandwiched between two STIs (see also Ref. \cite{Josephson-Witten} for another, closely related, example).  
If an external magnetic field is applied perpendicular to the interfaces, a flux line vortex lattice will 
arise and the magnetic flux will be nonzero. For STIs we generally have $\theta(L)=-\theta(0)=\theta$, with $\theta=\pi$  
for time-reversal (TR) invariant systems.   
Using $q=n(2e)$ (with $(2e)$ the Cooper pair charge) and considering a total flux $\Phi_B$ due to $N_v$ straight vortex lines, 
we obtain the total charge, 
\begin{equation}
\label{Eq:Witten-SC}
Q=e\left(2n-\frac{\theta N_v}{2\pi}\right).
\end{equation}

\subsection{The Hall conductivity and the Witten effect}

If there are no magnetic monopoles, we can derive the Hall conductivity from the current density obtained 
from Eq. (\ref{Eq:ED-L})., by assuming that there is an interface separating a topologically trivial insulator ($\theta=0$) from a 
topologically nontrivial one ($\theta\neq 0$).    
We then find a dissipationless Hall current \cite{Josephson-Witten} given by, 
\begin{equation}
{\bf j}_H=-\frac{e^2}{4\pi^2}(\nablab\theta\times{\bf E}).
\end{equation}
If we consider an electric field applied at the surface $z=0$, e.g., ${\bf E}=E\hat {\bf x}$, we obtain the transverse surface current 
\begin{equation}
i_y=-\frac{e^2E}{4\pi^2}\int_0^\infty dz\frac{d\theta}{dz}=\sigma_{xy}E,
\end{equation}
where the Hall conductivity \cite{Sitte2012},
\begin{equation}
\label{Eq:Hall}
\sigma_{xy}=\frac{e^2}{2\pi}\left(n-\frac{\theta}{2\pi}\right).
\end{equation}
We note the similarity between the expression for the charge, Eq. (\ref{Eq:Q-Witten}) and the one for the Hall conductivity, Eq. (\ref{Eq:Hall}). 
In the following, we will show that for the case of topological superconductors this is not a mere accident (note, however, that a 
superconductor has elementary charge $2e$ rather than $e$).  
 
This can already be seen by considering the very simple problem of a charged particle of mass $M$ constrained to 
move on a ring of radius $r$ and in the presence of a magnetic flux, $\Phi$. 
In this exactly solvable example it is easy to see that the current is 
given by,
\begin{equation}
j_n=-e\frac{dE_n}{d\Phi}=\frac{e^3}{2\pi Mr^2}\left(n-\frac{e\Phi}{2\pi}\right),
\end{equation}
where 
\begin{equation}
E_n(\Phi)=\frac{1}{2Mr^2}\left(n-\frac{e\Phi}{2\pi}\right)^2,~~~~~~~~n\in\mathbf{Z},
\end{equation}
are the exact energy eigenvalues.

\section{Compact Abelian Higgs model with axion term }
Since the Witten effect in axion electrodynamics arises either in the presence magnetic monopoles or vortices, a general Abelian Higgs 
model accounting for both topological defects is given by the Lagrangian written in imaginary time,
\begin{eqnarray}
\label{Eq:S-compact-Higgs}
\!\!\!\!{\cal L}=\frac{1}{4}{\cal F}_{\mu\nu}^2+\frac{ie^2\theta}{16\pi^2}{\cal F}_{\mu\nu}\tilde {\cal F}_{\mu\nu}
+ \frac{\rho^2}{2}(\partial_\mu\varphi+2eA_\mu)^2
+\frac{1}{2\rho_V^2}m_\mu^2.
\end{eqnarray}
Here, the field strength and its dual are given by \cite{Cardy}, 
\begin{equation}
\label{Eq:F}
{\cal F}_{\mu\nu}=F_{\mu\nu}+\frac{\pi}{e}\tilde M_{\mu\nu},~~~~~~
\end{equation}
and
\begin{equation}
\label{Eq:dual-F}
\tilde {\cal F}_{\mu\nu}=\tilde F_{\mu\nu}+\frac{\pi}{e}M_{\mu\nu}.
\end{equation}
with $F_{\mu\nu}=\partial_\mu A_\nu-\partial_\nu A_\mu$ and $\tilde F_{\mu\nu}=(1/2)\epsilon_{\mu\nu\lambda\rho}F_{\mu\nu}$. 
We also have that
$M_{\mu\nu}=\partial_\mu M_\nu-\partial_\nu M_\mu$ and $\tilde M_{\mu\nu}=(1/2)\epsilon_{\mu\nu\lambda\rho}M_{\mu\nu}$
where  
\begin{equation}
M_\mu(x)=\int d^4x'G(x-x')m_\mu(x'),  
\end{equation}  
with 
the Coulomb Green function $G(x)=1/(4\pi^2x^2)$. The field  
$m_\mu(x)$ is conserved and 
has the meaning of a magnetic monopole current. Thus, $M_\mu(x)$ is a monopole gauge field.  
We automatically have that $\partial_\mu M_\mu=0$ in view of the conservation of the 
monopole current; it follows that $\partial_\mu\tilde {\cal F}_{\mu\nu}=(\pi/e)m_\nu$, as expected. 
We will later see that the parameter $\rho_V$ emulates vortex stiffness. The way in which it appears in Eq. (\ref{Eq:S-compact-Higgs}), $\rho_V^{-1}$ represents the chemical potential of 
monopoles. As discussed in Ref. \cite{Cardy}, the field strength ${\cal F}_{\mu\nu}$  
is
a four-dimensional generalization of 
the superfluid velocity of two-dimensional superfluids \cite{NelKost}. 
The magnetic monopole contribution 
accounts for the compactness 
of the local $U(1)$ gauge group in the same way that point vortices in two-dimensional superfluids account for the periodicity of 
the phase of the superfluid wavefunction \cite{Cardy}.  This procedure allows one to incorporate the periodicity of 
lattice fields in a continuum field theory  approach where the fields become multivalued \cite{KleinertMultival}. 

In the absence of magnetic monopoles ($m_\mu=0$), Eq. (\ref{Eq:S-compact-Higgs}) describes a three-dimensional superconductor 
with a $\theta$-term, which can be realized via a heterostructure like the one shown in Fig. \ref{Fig:TI-SC-TI}. 
For $\theta=0$ and in the presence of monopoles, 
 the phase structure of Eq. (\ref{Eq:S-compact-Higgs}) has been discussed in the past 
using a lattice gauge theory formulation \cite{FS}, where it has been pointed out that the model with two units of 
charge features three phase rather than two. Indeed, for the case of one unit of charge the Higgs and confinement phases 
cannot be distinguished, differently of the case with two units of charge. The third phase in the problem is the Coulomb phase. 
There is a first-order phase transition between the Higgs and the confined phases \cite{Lattice-Higgs}.  
For $\theta\neq 0$ a first-order transition between the Higgs and the confinemed phases is still expected, but there are several such transitions, which 
are labeled by the integer monopole charge $m$ \cite{Cardy-theta,Cardy-Rabinovici}.   

Further insight into the theory  (\ref{Eq:S-compact-Higgs}) can be obtained by introducing an auxiliary field $h_\mu$ to rewrite it (see 
Appendix A) as, 
\begin{eqnarray}
\label{Eq:h}
{\cal L}'&=&\frac{1}{4}(F_{\mu\nu}^2+f_{\mu\nu}^2)+i\frac{e^2\theta}{16\pi^2}F_{\mu\nu} \tilde F_{\mu\nu}+im_\mu
\left(\frac{\pi}{e}h_\mu+\frac{e\theta}{4\pi}A_\mu\right)
\nonumber\\
&+&\frac{\rho^2}{2}(\partial_\mu\varphi+2eA_\mu)^2+\frac{1}{2\rho_V^2}m_\mu^2,
\end{eqnarray}
where $f_{\mu\nu}=\partial_\mu h_\nu-\partial_\nu h_\mu$. Physically, the gauge field $h_\mu$ 
accounts for the magnetic flux inside the vortex lines, akin to the London theory. Now, in order to integrate out the 
monopole gauge field subject to the constraint $\partial_\mu m_\mu=0$, we introduce a Lagrange multiplier $\varphi_V$ 
enforcing the constraint and perform the Gaussian integration over $m_\mu$ to obtain, 
\begin{eqnarray}
\label{Eq:dyonicSC}
{\cal L}&=&\frac{1}{4}(F_{\mu\nu}^2+f_{\mu\nu}^2)+i\frac{e^2\theta}{16\pi^2}F_{\mu\nu} \tilde F_{\mu\nu}
\nonumber\\
&+&\frac{\rho^2}{2}(\partial_\mu\varphi+2eA_\mu)^2+\frac{\rho_V^2}{2}\left(\partial_\mu\varphi_V+
\frac{\pi}{e}h_\mu+\frac{e\theta}{4\pi}A_\mu\right)^2.
\end{eqnarray}
The above Lagrangian indicates that $\varphi_V$ physically represents the phase of a vortex disorder field and that 
$\rho_V$ can be indeed be interpreted as a vortex stiffness.   
Due to the magnetoelectric (axionic) coupling, the vortex current couples directly to the vector potential with charge $e\theta/(4\pi)$. 

Despite similarities with the Ginzburg-Landau theory of three-dimensional topological superconductors discussed in Ref. \cite{Qi-Witten-Zhang}, 
Eq. (\ref{Eq:dyonicSC}) has a very different physical content. The theory of Ref. \cite{Qi-Witten-Zhang} features two 
superconducting order parameters coupled to the vector potential with charge $2e$, and $\theta$ is the phase difference between the 
phases of each order parameter. Furthermore, the gauge field $h_\mu$ is absent. 

The Lagrangian (\ref{Eq:dyonicSC}) for $\theta=0$ is a model for superconducting cosmic strings introduced by Witten quite some time ago 
\cite{Witten-SC-strings}.  Note that the presence of the $\theta$-term leads  to a fractionalization of the vortex string charge. 
Indeed, the vortex string charge is given by, 
\begin{eqnarray}
Q_V&=&S\int_Lds\left[\vphantom{\frac{e\theta}{4\pi}}2e\rho^2(\partial_t\varphi+2eA_0)\right.
\nonumber\\
&+&\left.\frac{e\theta}{4\pi}\rho_V^2\left(\partial_t\varphi_V+
\frac{\pi}{e}h_0+\frac{e\theta}{4\pi}A_0\right)\right],
\end{eqnarray}
where $S$ corresponds to a cross-sectional area of the string and the integral is along a path $L$ defined by the vortex line, which 
can also form closed loops in general. For $\theta=0$ the above equation reduces to the standard formula for the vortex charge.  

\section{Electromagnetic duality}

\subsection{Dual model}

In the absence of matter fields (i.e., $\rho=0$), the Lagrangian (\ref{Eq:S-compact-Higgs}) reduces to a compact Maxwell theory with 
an axion term. Note that for $\theta=0$ the two Higgs sectors in Eq. (\ref{Eq:dyonicSC}) decouple. The corresponding Higgs 
electrodynamics of vortices that is obtained in this way corresponds precisely to the model dual to the compact Maxwell theory in 
3+1 dimensions \cite{Peskin}.  For $\theta\neq 0$, the gauge field $A_\mu$ remains coupled to the vortex Higgs model when 
$\rho=0$. The compact Maxwell theory with an axion term has the same form as the Lagrangian for the electrodynamics of a 
topological insulator \cite{Qi-2008}, except that the latter case does not include magnetic monopoles. We may interpret the 
compact version of the axion electrodynamics of topological insulators as a model for topological interacting systems, like 
topological Mott insulators \cite{Senthil_Science-2014}. 

Up to the surface term, the Lagrangian (\ref{Eq:dyonicSC}) has an electromagnetic self-duality made transparent by a shift 
$h_\mu\to h_\mu-\frac{e^2\theta}{4\pi^2}A_\mu$, followed by the rescalings, $h_\mu\to 2eh_\mu$, $A_\mu\to (\pi/e)A_\mu$. Following these, 
the Lagrangian reads
\begin{eqnarray}
\label{Eq:dyonicSC-1}
{\cal L}&=&\frac{1}{4}\left[
\begin{array}{cc}
F_{\mu\nu} & f_{\mu\nu}
\end{array}
\right]
\left[
\begin{array}{cc}
\frac{\pi^2}{e ^2}+\frac{e^2\theta^2}{16\pi^2} & -\frac{e^2\theta}{2\pi}\\
\noalign{\medskip}
-\frac{e^2\theta}{2\pi} & 4e^2
\end{array}
\right]
\left[
\begin{array}{c}
F_{\mu\nu}\\
\noalign{\medskip}
f_{\mu\nu}
\end{array}
\right]+i\frac{\theta}{16}F_{\mu\nu} \tilde F_{\mu\nu}
\nonumber\\
&+&\frac{\rho^2}{2}(\partial_\mu\varphi+2\pi A_\mu)^2+\frac{\rho_V^2}{2}(\partial_\mu\varphi_V+
2\pi h_\mu)^2.
\end{eqnarray}  
From the above representation a duality first discussed in Ref. \cite{Cardy-theta} in the context of a $U(1)$ lattice gauge 
theory is obtained. It is given  by the transformations, 
\begin{equation}
\label{Eq:dual-e}
e'=\frac{1}{2}\sqrt{\frac{\pi^2}{e^2}+\frac{e^2\theta^2}{16\pi^2}},~~~~~~~
%
\theta'=-\frac{4\theta e^2}{\frac{\pi^2}{e^2}+\frac{e^2\theta^2}{16\pi^2}},
\end{equation}
with the field transformations, 
%
$A_\mu\to h_\mu,~h_\mu\to -A_\mu, ~ \{\varphi\to\varphi_V,~\rho\to\rho_V\},$ and $\{\varphi_V\to -\varphi,~\rho_V\to\rho\}$,
such that the Lagrangian is invariant up to the surface ($\theta$-) term, meaning that 
the Lagrangian is self-dual in the bulk. From Eq. (\ref{Eq:dyonicSC-1}),  
we realize that Eq.  (\ref{Eq:dual-e}) implies that the Dirac duality 
$e^2e'^2=\pi^2/4$ of the $\theta=0$ case is replaced by a matrix relation $M M'=(\pi^2/4)I$ when 
$\theta\neq 0$. Here, $M$ is the 
matrix appearing in Eq. (\ref{Eq:dyonicSC-1}), and $I$ is a $2\times 2$ identity matrix \cite{Note-1}. 
This electromagnetic duality emulates a symmetry, since  
broadly, dualities are unitary transformations that become symmetries at self-dual points  \cite{Zohar-PRL-2010}. 
In the context of topological states, symmetry related aspects of duality have been recently studied in terms of interacting 
Dirac fermions \cite{Wang-Senthil,Metlitski-Vishwanath,Witten-2015,Metlitski-2015,Fradkin-2015}.


We can integrate out $\varphi$ in Eq. (\ref{Eq:h}) by introducing a conserved charge current $j_\mu$ to obtain, 
\begin{eqnarray}
\label{Eq:transf-HS}
{\cal L}'&=&\frac{1}{4}(F_{\mu\nu}^2+f_{\mu\nu}^2)+i\frac{e^2\theta}{16\pi^2}F_{\mu\nu} \tilde F_{\mu\nu}+i\frac{\pi}{e}m_\mu h_\mu
\nonumber\\
&+&ie\left(2j_\mu+\frac{\theta}{4\pi}m_\mu\right)A_\mu
+\frac{1}{2\rho^2}j_\mu^2+\frac{1}{2\rho_V^2}m_\mu^2.
\end{eqnarray}
Due to the $\theta$-term, the  gauge field 
$A_\mu$ couples to both charge and monopole currents, implying that 
the physical current is 
\begin{equation}
eJ_\mu=2ej_\mu+\frac{e\theta}{4\pi}m_\mu.
\end{equation}
Thus, integrating $eJ_0$ over the volume yields, 
\begin{equation}
\label{Eq:q}
Q=e\left(2n+m\frac{\theta}{4\pi}\right),
\end{equation}
where $n, m\in\mathbb{Z}$, and we have assumed the normalizations,
\begin{equation}
\label{Eq:norm}
\int d^3x~j_0(x)=n,~~~~~~\int d^3x~m_0(x)=m,
\end{equation}
which shows that Eq. (\ref{Eq:q}) is yet another incarnation of the Witten effect. 
From Eq. (\ref{Eq:q}) we note the invariance $\theta\to\theta+8\pi$, $n\to n-m$ as a consequence of the periodicity 
of $\theta$ \cite{Cardy-theta}. 
Setting $j_\mu=0$ and $m_\mu=0$ reduces to the situation of a non-interacting topological insulator \cite{Qi-2008}.
We further distinguish here the following relevant special cases. 
When $j_\mu=0$ and $m_\mu\neq 0$, the theory describes 
an interacting topological insulator, since no charge is flowing and the gauge field is compact.  
If both $j_\mu$ and $m_\mu$ are nonzero, a polarized state 
of dipoles made of one electric and one magnetic charge, the so called dyon \cite{Schwinger-1969}, may form. If such 
polarized dyonic system is overall charge neutral, we obtain a diamagnetolectric rather than a dielectric type of insulator. 

If we integrate out $A_\mu$ and $h_\mu$ in Eq. (\ref{Eq:transf-HS}), we obtain the continuum version of the 
lattice dual model obtained by Cardy \cite{Cardy-theta} and Cardy and Rabinovici \cite{Cardy-Rabinovici},
\begin{eqnarray}
\label{Eq:dual-S}
\tilde S&=&\frac{1}{2}\left(\frac{\pi^2}{e^2}+\frac{e^2\theta^2}{16\pi^2}\right)\int d^4x\int d^4x'G(x-x')m_\mu(x)m_\mu(x')
\nonumber\\
&+&\frac{(2e)^2}{2}\int d^4x\int d^4x' G(x-x')j_\mu(x)j_\mu(x')
\nonumber\\
&+&\frac{e^2\theta}{2\pi}\int d^4x\int d^4x' G(x-x')j_\mu(x)m_\mu(x')
\nonumber\\
&+&\int d^4x\left(\frac{1}{2\rho^2}j_\mu^2+\frac{1}{2\rho_V^2}m_\mu^2\right),
\end{eqnarray}
apart from the local quadratic terms $j_\mu^2$ and $m_\mu^2$. 
The electromagnetic duality (\ref{Eq:dual-e}) holds once more,
provided that the replacements $m_\mu\to j_\mu$ and $j_\mu\to-m_\mu$ are made.  

Vortices and (superfluid) particles have large stiffnesses in 
the lattice formulation of Ref. \cite{Cardy-theta} or, equivalently, no chemical potentials for charge and magnetic currents.  
However, such local quadratic terms should be generated by short-distance fluctuations.  

Note that when $\rho\to 0$, corresponding to the regime of a compact Maxwell theory with an axion term, the currents $j_\mu$ are 
frozen to zero, and the dual action (\ref{Eq:dual-S}) becomes a vector Coulomb gas of magnetic monopole currents. 

\subsection{Renormalization aspects}

From the electromagnetic self-duality (\ref{Eq:dual-e}) we see that $e'^2\theta'=-e^2\theta$ and that, 
\begin{equation}
e'e=\frac{1}{2}\sqrt{\pi^2+\frac{e^4\theta^2}{16\pi^2}},
\end{equation}
must be invariant by renormalization, i.e., $e'_re_r=e'e$, where the subindex $r$ denotes renormalized counterparts. If $Z_A$ is the wavefunction renormalization for the field $A_\mu$, we obtain from the Ward identities the 
usual result, $e^2_r=Z_Ae^2$ following from gauge invariance. Thus, if $Z_h$ denotes the wavefunction renormalization for the field $h_\mu$, duality 
invariance immediately implies that $Z_AZ_h=1$. Therefore, if we use the Ward identities once more, we obtain,
\begin{equation}
\theta_r=\sqrt{Z_AZ_h}\theta=\theta,
\end{equation}
implying that $\theta$ does not renormalize. Thus, we have,  
\begin{equation}
e^2_r\theta_rF_{r,\mu\nu}\tilde F_{r,\mu\nu}=e^2\theta F_{\mu\nu}\tilde F_{\mu\nu},
\end{equation}
implying that the axion term is a renormalization invariant. This is consistent with the topological character of the axion term as a topological 
term. Indeed, since it does not depend on the metric, we expect it to be insensitive to scale transformations and therefore it must not change under 
renormalization.

\section{Boundary theory and duality at strong-coupling}

Since $F_{\mu\nu} \tilde F_{\mu\nu}=2\epsilon_{\mu\nu\lambda\rho}\partial_\mu(A_\nu\partial_\lambda A_\rho)$, 
the $\theta$-term yields a Chern-Simons (CS) term at the boundary. Thus, if we consider a system defined with a 
boundary at $z=0$,  the actual physics of the problem is described by a dimensionally reduced system in the strong-coupling limit. 
To see this we first write,
\begin{equation}
\frac{1}{4}F_{\mu\nu}^2=\frac{1}{2}(\epsilon_{\mu\nu\lambda}\partial_\nu A_\lambda)^2+\frac{1}{2}(\partial_zA_\mu-\partial_\mu A_z)^2,
\end{equation}
where now it is understood that the Greek indices on the RHS of the above equation 
refer to three-dimensional spacetime, $x_\parallel=(\tau,x,y)$, with a similar expression 
holding for $f_{\mu\nu}$. 
Thus, upon integrating out both $A_z$ and $h_z$, we 
obtain that the 
action associated to the Lagrangian (\ref{Eq:transf-HS}) can be written in the form,
\begin{widetext}
\begin{eqnarray}
\label{Eq:Sprime}
S'&=&\frac{1}{2}\int d ^4x\left[(\epsilon_{\mu\nu\lambda}\partial_\nu A_\lambda)^2
+(\epsilon_{\mu\nu\lambda}\partial_\nu h_\lambda)^2
+(\partial_zA_\mu)^2+(\partial_zh_\mu)^2+ieJ_\mu A_\mu+i\frac{\pi}{e}m_\mu h_\mu
+\frac{1}{2\rho^2}j_\mu^2+\frac{1}{2\rho_V^2}m_\mu^2\right]
\nonumber\\
&+&i\frac{e^2\theta}{8\pi^2}\int d ^3x_\parallel\epsilon_{\mu\nu\lambda}A_\mu\partial_\nu A_\lambda
+\frac{(2e)^2}{2}\int d ^3x_\parallel\int d ^3x_\parallel' G_{3D}(x_\parallel-x_\parallel')
j_z(x_\parallel)j_z(x_\parallel')
\nonumber\\
&+&\frac{1}{2}\left(\frac{\pi^2}{e^2}+\frac{e^2\theta^2}{16\pi^2}\right)\int d ^3x_\parallel\int d ^3x_\parallel' G_{3D}(x_\parallel-x_\parallel')
m_z(x_\parallel)m_z(x_\parallel')
+\frac{e^2\theta}{2\pi}\int d ^3x_\parallel\int d ^3x_\parallel' G_{3D}(x_\parallel-x_\parallel')
j_z(x_\parallel)m_z(x_\parallel')
\end{eqnarray}
where $G_{3D}(x_\parallel-x_\parallel')=1/(4\pi|x_\parallel-x_\parallel'|)$ and, 
\begin{equation}
j_z(x_\parallel)=\int_{-\infty}^{\infty}dzj_z(x_\parallel,z),~~~~~~~~~~
m_z(x_\parallel)=\int_{-\infty}^{\infty}dzm_z(x_\parallel,z),
\end{equation}
\end{widetext}
and we have used that $\theta(z)=\theta$ for $z\geq 0$, vanishing otherwise.  
The second and third lines of Eq. (\ref{Eq:Sprime}) contain only surface modes, while the bulk still contributes 
in the first line. 

An interesting  limiting case where the boundary theory decouples from the bulk is obtained by letting $e^2\to\infty$. By 
rescaling $A_\mu\to e^{-1}A_\mu$ and 
$h_\mu\to e^{-1}h_\mu$ in Eq. (\ref{Eq:transf-HS}), the action for $e^2\to\infty$ becomes, 
\begin{eqnarray}
\label{Eq:S-infinity}
&&S_\infty =i\frac{\theta}{8\pi^2}\int d ^3x_\parallel\epsilon_{\mu\nu\lambda}A_\mu\partial_\nu A_\lambda
\nonumber\\
&+&\int d^4x\left[
i\left(2j_\mu+\frac{\theta}{4\pi}m_\mu\right)A_\mu
+\frac{1}{2\rho^2}j_\mu^2+\frac{1}{2\rho_V^2}m_\mu^2\right].
\end{eqnarray} 
Because there is no Maxwell term in $S_\infty$, we have that $J_\mu=0$ in the bulk and the currents exist only on the surface, i.e., we 
have an insulating bulk.  
From Eq. (\ref{Eq:Sprime}) we also see that both $j_z$ and $m_z$ are constrained to vanish in the limit $e^2\to\infty$.  
Since $J_\mu$ vanishes in the bulk, 
Eq. (\ref{Eq:q}) implies that $\theta/(8\pi)=-n/m$, $m\neq 0$.  This result is consistent with Cardy's discussion \cite{Cardy-theta} 
of the phase structure of the lattice model, although the boundary theory has not been considered in Ref. \cite{Cardy-theta}. 
There the critical point is attained at values $\theta/(2\pi)=-n/m$ (note the factor $2\pi$ instead of $8\pi$, which arises in our 
case because the charge of our bosons is $2e$), when the bare coupling becomes infinitely large. 

Note that locking $\theta$ to $-8\pi n/m$ in the strong-coupling regime implies that 
$\theta$ {\it cannot be smoothly connected to zero}, corresponding to a situation similar to the one encountered recently 
\cite{Nogueira-Sudbo-Eremin}  in the renormalization group 
analysis of a three-dimensional topological superconductor of the type studied in Ref. \cite{Qi-Witten-Zhang}. 
 In the following we will elaborate further on this regime by means of the duality transformation. 

A subtle aspect of the boundary theory following from Eq. (\ref{Eq:S-infinity}) is uncovered when performing the Gaussian integral over $A_\mu$. 
Integrating out  $A_\mu$ at the boundary leads to the effective Lagrangian at strong coupling ($J_\mu\neq 0$ at the boundary),
\begin{equation}
\label{Eq:J-CS}
\tilde {\cal L}_\infty=i\frac{2\pi^2}{\theta}\epsilon_{\mu\nu\lambda}J_\mu\frac{\partial_\nu}{\partial^2} J_\lambda 
+\frac{1}{2\rho^2}j_\mu^2+\frac{1}{2\rho_V^2}m_\mu^2.
\end{equation}
Solving the current conservation constraints yields, $j_\mu=\epsilon_{\mu\nu\lambda}\partial_\nu a_\lambda$ and 
$m_\mu=\epsilon_{\mu\nu\lambda}\partial_\nu b_\lambda$. Therefore, 
\begin{eqnarray}
\label{Eq:CS-strong}
\tilde {\cal L}_\infty&=&\frac{1}{2\rho^2}(\epsilon_{\mu\nu\lambda}\partial_\nu a_\lambda)^2+
\frac{1}{2\rho_V^2}(\epsilon_{\mu\nu\lambda}\partial_\nu b_\lambda)^2
\nonumber\\ 
&+&i\frac{2\pi^2}{\theta}\epsilon_{\mu\nu\lambda}\left(2a_\mu+\frac{\theta}{4\pi}b_\mu\right)
\partial_\nu \left(2a_\lambda+\frac{\theta}{4\pi}b_\lambda\right).
\end{eqnarray}
If we define a two component gauge field $(h_{I\mu})=(a_\mu,b_\mu)$, we can rewrite the above Lagrangian in the form, 
\begin{equation}
\label{Eq:K-matrix-L}
\tilde {\cal L}_\infty=\frac{1}{2}\sum_I\frac{1}{\rho_I}(\epsilon_{\mu\nu\lambda}\partial_\nu h_{I\lambda})^2
+i\pi\epsilon_{\mu\nu\lambda}\sum_{I,J}K_{IJ}h_{I\mu}\partial_\nu h_{J\lambda},
\end{equation}
where $\rho_1=\rho$ and $\rho_2=\rho_V$, and $K_{IJ}$ are the elements of the matrix,
\begin{equation}
K=\left[
\begin{array}{cc}
-m/n & 1 \\
\noalign{\medskip}
1 &-n/m
\end{array}
\right],
\end{equation}
The result of a matrix CS term is reminiscent from effective theories for the fractional quantum Hall state \cite{Wen-Zee}. Actually, 
our system is rather an anyon superfluid, as $\det K=0$, implying the existence of a gapless mode.   Note, however, that the entries 
of the matrix $K$ are not necessarily integers in this case. 

The Lagrangian (\ref{Eq:CS-strong}) describes a free theory leading us to conclude that the strongly coupled theory at the boundary 
is non-interacting. However, this is an example where standard continuum manipulations yield an 
incorrect result. The Lagrangian (\ref{Eq:CS-strong}) is actually incomplete, as an analysis made in the lattice will now demonstrate. 
The difficulty lies on the fact that solving the current conservation constraint in the continuum formulation misses in some cases 
the periodic character of phase variables that underly the current conservation itself. 
There is, in fact, a discrete 
periodicity in the current that cannot always be properly captured with a field-theoretical analysis performed directly 
in the continuum.  

The lattice boundary theory associated with the bulk action (\ref{Eq:S-infinity}) is
\begin{eqnarray}
\label{Eq:S-inf-lattice}
S_{\infty}^b&=&\sum_l\left[i\frac{\theta}{8\pi^2}\epsilon_{\mu\nu\lambda}A_{l\mu}\Delta_\nu A_{l\lambda}
+i\left(2j_{l\mu}+\frac{\theta}{4\pi}m_{l\mu}\right)A_{l\mu}\right.
\nonumber\\
&+&\left.\frac{1}{2\rho^2}j_{l\mu}^2+\frac{1}{2\rho_V^2}m_{l\mu}^2\right],
\end{eqnarray}
where the lattice derivative is defined in a standard way as $\Delta_\mu f_l=f_{l+1}-f_l$ (with unit lattice spacing).
The currents $j_{l\mu}$ and $m_{l\mu}$ are now integer valued lattice fields, making the normalization 
 superfluous. Thus, the partition function
\begin{equation}
\label{Eq:partition}
Z_\infty^b=\sum_{\{j_{l\mu}\}}\sum_{\{m_{l\mu}\}}\delta_{\Delta_\mu j_{l\mu},0}\delta_{\Delta_\mu m_{l\mu},0} 
\int_{-\infty}^\infty \left[\prod_j dA_{j\mu}\right]e^{-S_\infty^b},
\end{equation}
with the current conservation constraints being enforced by Kronecker deltas. 
Using the integral representation of the Kronecker deltas, 
\begin{equation}
\delta_{\Delta_\mu n_{l\mu},0}=\int_{0}^{2\pi}\frac{d\varphi_{l}}{2\pi}e^{i\varphi_{l}\Delta_\mu n_{l\mu}}, 
\end{equation}
\begin{equation}
\delta_{\Delta_\mu s_{l\mu},0}=\int_{0}^{2\pi}\frac{d\varphi_{Vl}}{2\pi}e^{i\varphi_{Vl}\Delta_\mu s_{l\mu}}, 
\end{equation}
in Eq. (\ref{Eq:partition}), 
and applying once more the Poisson formula \cite{Note-3}, 
\begin{eqnarray}
\label{Eq:S-inf-lattice-1}
S_{\infty}^b&=&\sum_l\left[i\frac{\theta}{8\pi^2}\epsilon_{\mu\nu\lambda}A_{l\mu}\Delta_\nu A_{l\lambda}
+\frac{\rho^2}{2}(\Delta_\mu\varphi_{l}-2\pi p_{l\mu}-2 A_{l\mu})^2\right. 
\nonumber\\
&+&\left. \frac{\rho_V^2}{2}\left(\Delta_\mu\varphi_{Vl}-2\pi q_{l\mu}-\frac{\theta}{4\pi}A_{l\mu}\right)^2
\right],
\end{eqnarray}
with another set of integer fields, $p_{l\mu}$ and $q_{l\mu}$. 

Integrating over $A_{l\mu}$ yields a lattice version of Eq. (\ref{Eq:J-CS}) 
where the currents are integer fields. Solving the current conservation constraints yields  
integer-valued gauge fields, $j_{l\mu}=\epsilon_{\mu\nu\lambda}\Delta_\nu N_{l\lambda}$ and 
$m_{l\mu}=\epsilon_{\mu\nu\lambda}\Delta_\nu M_{l\lambda}$. This point is the key to understand why Eq. (\ref{Eq:CS-strong}) is not quite 
correct. The corresponding lattice action has the same form as Eq. (\ref{Eq:CS-strong}), 
but with integer-valued gauge fields. Introducing real-valued lattice gauge fields via the Poisson formula \cite{Note-2}, we obtain 
\begin{eqnarray}
\label{Eq:CS-strong-1}
\tilde {S}_\infty^b&=&\sum_l\left[\frac{1}{2\rho^2}(\epsilon_{\mu\nu\lambda}\Delta_\nu a_{l\lambda})^2+
\frac{1}{2\rho_V^2}(\epsilon_{\mu\nu\lambda}\Delta_\nu b_{l\lambda})^2\right.
\nonumber\\ 
&+&i\frac{2\pi^2}{\theta}\epsilon_{\mu\nu\lambda}\left(2a_{l\mu}+\frac{\theta}{4\pi}b_{l\mu}\right)
\Delta_\nu \left(2a_{l\lambda}+\frac{\theta}{4\pi}b_{l\lambda}\right)
\nonumber\\
&-&\left. 2\pi in_{l\mu}a_{l\mu}-2\pi is_{l\mu}b_{l\mu}
\right],
\end{eqnarray}
where $n_{l\mu}$ and $s_{l\mu}$ are integer fields representing conserved currents, which in this case is a consequence 
of gauge invariance. In contrast to Eq. (\ref{Eq:CS-strong}), due to the coupling of the gauge fields to the currents,  Eq. (\ref{Eq:CS-strong-1}) does not yield a 
free quadratic theory.

The action $\tilde {S}_\infty^b$ corresponds to the boundary dual of the action ${S}_\infty^b$. 
Besides realizing that the theory given by Eq. (\ref{Eq:J-CS}) is actually not free, the dual transformation above shows that $\rho$ and $\rho_V$ of the action ${S}_\infty^b$ become the dielectric constants (or gauge couplings) in 
the dual action $\tilde {S}_\infty^b$. While ${S}_\infty^b$ is strongly coupled, $\tilde {S}_\infty^b$ is not. 
This allows us to find a regime where the boundary theory becomes self-dual. The self-dual regime is expediently explored 
using the actions of Eqs. (\ref{Eq:S-inf-lattice},\ref{Eq:CS-strong-1}). In Eq. (\ref{Eq:S-inf-lattice}), 
$m_{l\mu}$ vanishes when $\rho_V\to 0$. Similarly, in Eq. (\ref{Eq:CS-strong-1}), $b_{l\mu}$ can be gauged away in this limit. Thus, by assuming the limit $\rho_V\to 0$ 
and rescaling $A_{l\mu}\to \pi A_{l\mu}$, we obtain the self-duality of the actions  (\ref{Eq:S-inf-lattice}) and (\ref{Eq:CS-strong-1}) at 
$\rho\to\infty$, provided $\theta/(8\pi)=\pm 1$, in which case the actions become precisely equivalent.

\section{Possible generalizations}

It is in principle possible to connect the compact Maxwell theory discussed in this paper to quantum spin models exhibiting an emergent $U(1)$ symmetry, 
like for example, those models described by the theory of deconfined quantum critical points \cite{Senthil-2004}. 
In order to put in perspective the types of bosonic topological states we are looking for in terms of spins models, we start 
by recalling some properties of deconfined critical points in 2+1 dimensions that are useful in this paper. 
We first consider  a version of the  Faddeev-Skyrme model \cite{FS-model,Babaev}  as discussed similarly in 
Ref. \cite{Nogueira-Sudbo-2012},   
\begin{equation}
\label{FSB}
{\cal L}=\frac{1}{2g}(\partial_\mu{\bf n})^2+\frac{1}{2e^2}[\epsilon_{\mu\nu\lambda}\partial_\nu c_\lambda+\epsilon_{\mu\nu\lambda}{\bf n}
\cdot(\partial_\nu{\bf n}\times\partial_\lambda{\bf n})]^2,
\end{equation} 
where ${\bf n}^2=1$ and $c_\mu$ is a non-compact $U(1)$ gauge field.  
The strongly coupled regime $g\to \infty$ describes a nontrivial paramagnetic phase where  
the Lagrangian (\ref{FSB}) becomes a compact Maxwell theory. This can be shown by using 't Hooft's  construction \cite{tHooft-1974} 
of an Abelian gauge field from a non-Abelian one. Indeed, we can write $F_{\mu\nu}=
{\bf n}\cdot{\bf F}_{\mu\nu}=\partial_\mu c_\nu-\partial_\nu c_\mu+{\bf n}\cdot(\partial_\mu{\bf n}\times\partial_\nu{\bf n})$, 
where ${\bf F}_{\mu\nu}=\partial_\mu{\bf J}_\nu-\partial_\nu{\bf J}_\mu-{\bf J}_\mu\times{\bf J}_\nu$ 
is a non-Abelian field strength associated with the $O(3)$ gauge field, ${\bf J}_\mu={\bf n}c_\mu+{\bf n}\times\partial_\mu{\bf n}$. Since, 
\begin{equation}
\label{hedgehog-af}
Q=\frac{1}{8\pi}\oint_{S}dS_\mu\epsilon_{\mu\nu\lambda}{\bf n}\cdot(\partial_\nu{\bf n}
\times\partial_\lambda{\bf n}),
\end{equation}
where $Q\in\mathbb{Z}$, 
%
the $g\to\infty$ limit of the theory dualizes to a sine-Gordon theory with $\pi/2$ 
periodicity, rather than the usual $2\pi$ one of Polyakov's compact Maxwell theory in 2+1 dimensions \cite{Polyakov}.    
Physically the $\pi/2$ periodicity represents the $\pi/2$ rotations mapping a VBS state into another one \cite{Senthil-2004}. 
Since the sine-Gordon model in 2+1 dimensions is always gapped, there is no phase transition occurring in the system. This gap leads 
to a finite string tension between spinons and anti-spinons in the original model, which impedes a deconfinement to occur.  
Since $[\epsilon_{\mu\nu\lambda}{\bf n}\cdot(\partial_\nu{\bf n}\times\partial_\lambda{\bf n})]^2=(\partial_\mu{\bf n}\times\partial_\nu{\bf n})^2$, 
and ${\bf n}$ is the direction of the spin, a lattice model associated to the compact Maxwell term would 
automatically include four-spin 
interactions between singlet bonds, similarly to the so called $J-Q$ model \cite{Sandvik_2007}. The limit $g\to\infty$ corresponds 
to the case where the four-spin singlet bond interaction dominates the physics.   

A topologically nontrivial theory in 2+1 dimensions can be obtained by taking 't Hooft's construction one step further to 
add into the Lagrangian (\ref{FSB}) the (non-abelian) CS term,
\begin{eqnarray}
\label{Eq:CS}
{\cal L}_{\rm CS}&=&i\frac{\theta}{16\pi^2}\epsilon_{\mu\nu\lambda}\left[{\bf J}_\mu\cdot\partial_\nu{\bf J}_\lambda
+\frac{1}{3}{\bf J}_\mu\cdot({\bf J}_\nu\times{\bf J}_\lambda)\right]
\nonumber\\
&=&i\frac{\theta}{16\pi^2}\left[\epsilon_{\mu\nu\lambda}c_\mu\partial_\nu c_\lambda-\frac{2}{3}c_\mu 
\epsilon_{\mu\nu\lambda}{\bf n}\cdot(\partial_\nu{\bf n}\times\partial_\lambda{\bf n}) \right]. 
\end{eqnarray}
One way to realize the above CS contribution in models of quantum criticality in 2+1 dimensions is to assume a physical 
situation where the quantum phase transition occurs on the surface of a (3+1)-dimensional system.  In this case, the CS term 
arises from a so called $\theta$-term in the action of a (3+1)-dimensional theory, which has the well-known form, 
\begin{equation}
{\cal S}_\theta=i\frac{\theta}{32\pi^2}\int d^4x\epsilon_{\mu\nu\lambda\rho}{\bf F}_{\mu\nu}\cdot{\bf F}_{\lambda\rho}
=i\frac{\theta}{32\pi^2}\int d^4x\partial_\mu K_\mu,
\end{equation}
where $K_\mu=2\epsilon_{\mu\nu\lambda\rho}[{\bf J}_\nu\cdot\partial_\lambda{\bf J}_\rho
+(1/3){\bf J}_\nu\cdot({\bf J}_\lambda\times{\bf J}_\rho)]$ is the CS current.  Again, it is possible to define a compact abelian 
$\theta$-term from the non-abelian one. Within this point of view, 
a topological interacting state of matter in three dimensions 
mimics the electrodynamics of topological band insulators \cite{Qi-2008}, where   
a fluctuating field associated to topological defects leads to an emergent compact $U(1) $ symmetry. 

\section{Conclusions}

We have constructed and exploited the dualities of a compact abelian Higgs model with a topological axion term and shown that it is 
equivalent to a topological, non-compact, abelian Higgs model having two Higgs and two gauge fields, akin to the model for 
superconducting vortex strings, but with a topological term.  In other words, we have established the equivalence between a 
topological theory having bosonic particles coupled to monopoles in a gauge invariant way and a 
topological theory having bosonic particles and vortices. 
This equivalence allows us to better understand how the Witten effect also applies to a system having vortex lines and no monopoles: 
the two versions of the Witten effect are simply dual to each other. 

The duality is particularly interesting when the topological field theory system has  a boundary, like the cases that typically arise in 
topological condensed matter states of matter \cite{Hasan-Kane-RMP,Zhang-RMP-2011}.  In particular, we have shown that in the 
strongly interacting regime $\theta=-8\pi n/m$, with $n$ and $m$ being integers ($m\neq 0$). The same quantization appears 
at infinite coupling critical point of the bulk lattice theory, as previously demonstrated via symmetry arguments involving 
modular transformations \cite{Cardy-theta}. The strong-coupling boundary theory features two gauge fields and a mutual CS term. We have 
shown that its dual exactly corresponds to a two-scalar field  Higgs model  coupled to a single gauge field whose dynamics is governed 
by the CS term, with no Maxwell term in the Lagrangian. Interestingly, the scalar field associated to the vortices provides a charge that is 
topologically induced, being just given by $\theta/(4\pi)=-2n/m$.  

\acknowledgments

F.S.N. and J.v.d.B. would like to thank the Collaborative Research Center SFB 1143 ``Correlated Magnetism: From Frustration to Topology" for the financial support. ZN acknowledges partial support from NSF (CMMT) under grant number 1411229. 

\appendix
\begin{widetext}
\section{Derivation of Eq. (8) from Eq. (7)}

We have, 
\begin{equation}
{\cal F}_{\mu\nu}^2=F_{\mu\nu}^2+\frac{2\pi}{e}\epsilon_{\mu\nu\alpha\beta}F_{\mu\nu}\partial_\alpha
\int d^4x'G(x-x')m_\beta(x')+\left(\frac{\pi}{e}\right)^2\epsilon_{\mu\nu\lambda\rho}\epsilon_{\mu\nu\alpha\beta}
\int d^4x'\int d^4x''\partial_\lambda G(x-x')\partial_\alpha G(x-x'')m_\beta(x')m_\rho(x''). 
\end{equation}
It turns out that the second term in the above equation vanishes, while for the last term we use, 
\begin{equation}
\epsilon_{\mu\nu\lambda\rho}\epsilon_{\mu\nu\alpha\beta}=2(\delta_{\lambda\alpha} \delta_{\rho\beta} -
\delta_{\lambda\beta} \delta_{\rho\alpha} ). 
\end{equation}
Thus, in the action we obtain a contribution,
\begin{equation}
2\int d^4x\int d^4x'\int d^4x''G(x-x')\underbrace{[-\partial^2G(x-x'')]}_{=\delta^4(x''-x)}m_\rho(x)m_\rho(x''),
\end{equation}
where we have used integration by parts along with $\partial_\lambda G(x-x')=-\partial_\lambda^{'}G(x-x')$, for the term 
proportional to $\delta_{\lambda\beta} \delta_{\rho\alpha}$, and $\partial_\lambda^{'} m_\lambda(x')=0$. Therefore, 
the Maxwell term in the action reads, 
\begin{equation}
S_{\rm Maxwell}=\frac{1}{4}\int d^4xF_{\mu\nu}^2+\frac{\pi^2}{2e^2}\int d^4x\int d^4x'G(x-x')m_\rho(x)m_\rho(x'). 
\end{equation}
In view of the constraint $\partial_\mu m_\mu=0$, we can introduce an auxiliary field to rewrite the above equation in the form, 
\begin{equation}
S_{\rm Maxwell}=\frac{1}{4}\int d^4x\left[(F_{\mu\nu}^2+f_{\mu\nu}^2)+i\frac{\pi}{e}h_\mu m_\mu\right],
\end{equation}
where $f_{\mu\nu}=\partial_\mu h_\nu-\partial_\nu h_\mu$. 

For the $\theta$-term we have, 
\begin{eqnarray}
\epsilon_{\mu\nu\lambda\rho}{\cal F}_{\mu\nu}{\cal F}_{\lambda\rho}&=&\epsilon_{\mu\nu\lambda\rho}{F}_{\mu\nu}{F}_{\lambda\rho}+\frac{2\pi}{e}
\epsilon_{\mu\nu\lambda\rho}\epsilon_{\lambda\rho\alpha\beta}F_{\mu\nu}\partial_{\alpha}\int d^4x'G(x-x')m_\beta(x')
\nonumber\\
&+&\left(\frac{\pi}{e}\right)^2\epsilon_{\mu\nu\lambda\rho}\epsilon_{\mu\nu\alpha\beta}\epsilon_{\lambda\rho\gamma\delta}
\int d^4x'\int d^4x''\partial_\alpha G(x-x')\partial_\gamma G(x-x'')m_\beta(x')m_\delta(x''). 
\end{eqnarray}
Now, we have to use, 
\begin{equation}
\epsilon_{\mu\nu\lambda\rho}\epsilon_{\lambda\rho\alpha\beta}=2(\delta_{\mu\alpha}\delta_{\nu\beta}-\delta_{\mu\beta}\delta_{\nu\alpha}),
\end{equation}
and, similarly, 
\begin{equation}
\epsilon_{\mu\nu\lambda\rho}\epsilon_{\lambda\rho\gamma\delta}=2(\delta_{\mu\gamma}\delta_{\nu\delta}-
\delta_{\mu\delta}\delta_{\nu\gamma}).
\end{equation}
Thus, 
\begin{eqnarray}
S_{\rm axion}&=&i\frac{e^2\theta}{32\pi^2}\epsilon_{\mu\nu\lambda\rho}\int d^4x {\cal F}_{\mu\nu}{\cal F}_{\lambda\rho}
=i\frac{e^2\theta}{32\pi^2}\left[\int d^4x\epsilon_{\mu\nu\lambda\rho}{F}_{\mu\nu}{F}_{\lambda\rho}
+\frac{4\pi}{e}(\delta_{\mu\alpha}\delta_{\nu\beta}-\delta_{\mu\beta}\delta_{\nu\alpha})\int d^4x\int d^4x'F_{\mu\nu}(x)
\partial_\alpha G(x-x')m_\beta(x')\right.\nonumber\\
&+&\left. 2\left(\frac{\pi}{e}\right)^2(\delta_{\mu\gamma}\delta_{\nu\delta}-
\delta_{\mu\delta}\delta_{\nu\gamma})\epsilon_{\mu\nu\alpha\beta}\int d^4x\int d^4x'\int d^4x''
\partial_\alpha G(x-x')\partial_\gamma G(x-x'')m_\beta(x')m_\delta(x'')
\right]. 
\end{eqnarray}
Integration by parts produces, 
\begin{equation}
\int d^4x\int d^4x'\partial_\mu A_\nu(x)\partial_\mu G(x-x')m_\nu(x')=\int d^4xA_\nu(x)m_\nu(x), 
\end{equation}
such that, 
\begin{eqnarray}
S_{\rm axion}&=&i\frac{e^2\theta}{32\pi^2}\epsilon_{\mu\nu\lambda\rho}\int d^4x {\cal F}_{\mu\nu}{\cal F}_{\lambda\rho}
=i\frac{e^2\theta}{32\pi^2}\left\{\int d^4x\epsilon_{\mu\nu\lambda\rho}{F}_{\mu\nu}{F}_{\lambda\rho}
+\frac{8\pi}{e}\int d^4xA_\nu(x)m_\nu(x)-\frac{8\pi}{e}\int d^4x\int d^4x'\partial_\mu A_\mu(x)\partial_\nu G(x-x')m_\mu(x')
\right.\nonumber\\
&+&\left. \frac{2\pi^2}{e^2}\epsilon_{\mu\nu\alpha\beta}\int d^4x\int d^4x'\int d^4x''[
\partial_\alpha G(x-x')\partial_\mu G(x-x'')m_\beta(x')m_\nu(x'')-
\partial_\alpha G(x-x')\partial_\nu G(x-x'')m_\beta(x')m_\mu(x'')]
\right\}
\nonumber\\
&=&i\frac{e^2\theta}{32\pi^2}\left[\int d^4x\epsilon_{\mu\nu\lambda\rho}{F}_{\mu\nu}{F}_{\lambda\rho}
+\frac{8\pi}{e}\int d^4xA_\nu(x)m_\nu(x)+\frac{8\pi}{e}\int d^4x\int d^4x' A_\nu(x)\partial_\nu
\underbrace{\partial_\mu G(x-x')}_{=-\partial_\mu^{'}G(x-x')} m_\mu(x')\right.\nonumber\\
&+&\left. \frac{4\pi^2}{e^2}\epsilon_{\mu\nu\alpha\beta}\int d^4x\int d^4x'\int d^4x''
\partial_\alpha G(x-x')\partial_\mu G(x-x'')m_\beta(x')m_\nu(x'')
\right].
\end{eqnarray}
Therefore, after some final algebraic manipulations, we obtain that the sum of the Maxwell and axion actions yields, 
\begin{equation}
S_{\rm Maxwell}+S_{\rm axion}=\int d^4x\left[\frac{1}{4}(F_{\mu\nu}^2+f_{\mu\nu}^2)+i\frac{e^2\theta}{32}
\epsilon_{\mu\nu\lambda\rho}F_{\mu\nu}F_{\lambda\rho}+\left(\frac{e\theta}{4\pi}A_\mu+\frac{\pi}{e}h_\mu\right)m_\mu\right].
\end{equation}
\end{widetext}


\begin{thebibliography}{100}
\bibitem{Note-1} Note that $e^2e'^2=\pi^2/4$ is obtained instead of the more familiar form, $e^2e'^2=4\pi^2$, since 
the charge of the Cooper pair is $2e$.
\bibitem{Note-2} The Poisson summation formula is given by 
$$
\sum_{n=-\infty}^{\infty}f(n)=\sum_{m=-\infty}^{\infty}
\int_{-\infty}^{\infty}\frac{dx}{2\pi}f(x)e^{i2\pi mx}.
$$
\bibitem{Note-3} The form of the Poisson summation formula that is useful in this case corresponds to 
the identity, 
$$
\sum_{n=-\infty}^{\infty}e ^{-\frac{\epsilon}{2}n^2+ixn}=\sqrt{\frac{2\pi}{\epsilon}}\sum_{m=-\infty}^{\infty} 
e^{-\frac{1}{2\epsilon}(x-2\pi m)^2}.
$$
\end{thebibliography}
\end{document}